\documentclass[12pt]{iopart}
\usepackage{graphicx}
\begin{document}
\title[Guided vortex motion in a ferromagnet-superconductor bilayer]{Guided vortex motion in a ferromagnet-superconductor bilayer}
\author{P Gier{\l}owski$^1$,  I Abaloszewa$^1$, P Przys{\l}upski$^1$, and G Jung$^{1,2}$}
\address{$^1$Institute of Physics, Polish Academy of Sciences, 02-668 Warszawa, Al. Lotnik{\'o}w 32/46, Poland}
\address{$^2$Department of Physics, Ben Gurion University of the Negev, P.O.B. 653, 84105 Beer Sheva, Israel}

\ead{piotr.gierlowski@ifpan.edu.pl}

\begin{abstract}
Vortex motion in ferromagnetic-superconducting bilayers constituted by thin {\rm YBa$_2$Cu$_3$O$_{7-\delta}$} film deposited on the top of thin manganite {\rm La$_{0.67}$Sr$_{0.33}$MnO$_3$} film has been investigated. Both films were deposited on uniformly twinned single-crystalline substrates and patterned into strips directed at different angles to the twin lines pattern. These conditions lead to the guided vortex motion at different angles to the periodic pinning potential. It has been established that critical current and flux flow resistance in such an anisotropic pinning landscape-scale follow the elliptic dependence on the angle of the direction of the strip. For vortex flow at the directions close to the perpendicular to magnetic domain walls, signatures of the coherence in the vortex motion have been observed.
\end{abstract}

\vspace{2pc}
 \\

\maketitle

\section{Introduction}

\noindent
Magnetic interactions between ferromagnetic patterns layered on top of superconducting films and vortices lead to the magnetic pinning and appearance of unusual vortex arrangements \cite{Erdin}.  Such structures attract a lot of attention as they open new perspectives for the development of spintronics devices \cite{Wolf, Buzdin}.  The behavior of vortices in hybrid ferromagnetic superconductor multilayers depends strongly on the interaction of vortices with the pinning potential created by magnetic domains in the ferromagnetic layer \cite{Lyuksyutov}. Magnetic domains and domain walls in ferromagnetic systems affect the vortex dynamics in very different ways, depending on the shape and strength of the ferromagnetic pattern \cite{Goa}.\\

\noindent
Colossal magnetoresistance manganite perovskites, with a high degree of spin polarization in the ferromagnetic state and very good matching of crystalline structure with high-$T_c\,$ cuprate superconductors, are particularly well suited for growing the ferromagnet-superconductor bilayers \cite{Przyslupski,Holden}.
Magnetic domain patterns of a manganite film deposited on a twinned single crystalline substrate repeat the twin boundaries network arrangement. Deposition of a manganite film on the crystalline substrate with twin boundaries results in the appearance of manganite magnetic domains with well-defined in-plane magnetization, separated by pinned, out-of-plane magnetic structures in the film \cite{Laviano_2005,Laviano_2007}. Substrates possessing sufficiently large areas of uniformly twinned structures enable thus obtaining manganite films with regular strip patterns of magnetic domains.\\

\noindent
By assembling a ferromagnetic-superconducting bilayer structure of ferromagnetic manganite film with Curie temperature much higher than the temperature of superconducting transition in the superconducting film one ensures that the magnetic fingerprints are strongly embedded in the superconducting film even at zero applied magnetic fields. For regular and periodic stripe domain patterns, vortices moving along the direction of domain walls are expected to be channeled into channels of easy vortex flow, while those flowing in the direction perpendicular to the domain walls will move in a periodic pinning environment. In such situation one may expect to observe features of the coherent vortex motion and, possibly, the appearance of quasi-Josephson effects \cite{Zhu,Yuzhelevski,Shapiro}. This may constitute a basis for the development of cryoelectronic devices based on current-induced and driven vortex motion.\\

\noindent
In this paper, we report on the investigations of motion of current-induced vortices, in zero applied magnetic fields, in the ferromagnet-superconductor bilayers with periodic magnetic domain structures. In particular, by pattering the film into stripes directed at different angles  to the direction of the stripe magnetic domain structure we enforce vortex motion at the predetermined angles with respect to the direction of the stripe domains. The study focuses on the effects of magnetic domains interaction with current-driven vortices in zero applied magnetic field in the conditions foreseen for development of future vortex cryoelectronics devices.\\

\begin{figure}[h]
    \centering
    \includegraphics[width=8cm]{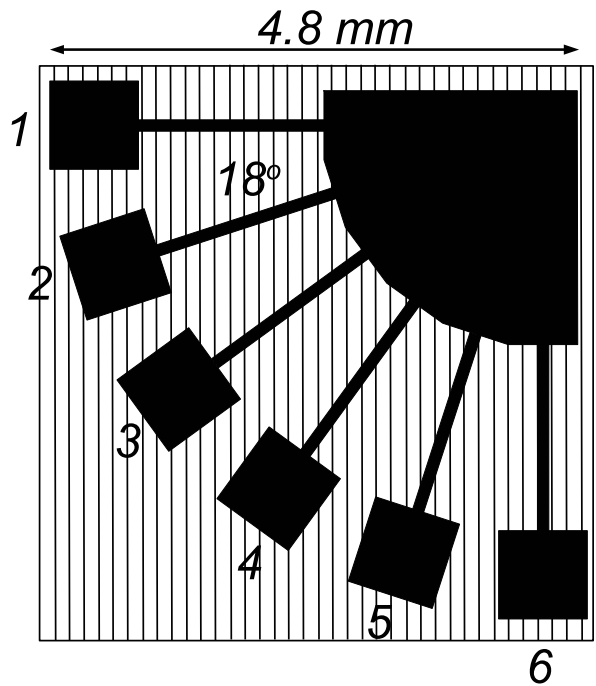}
    \caption{Pattern of our thin film structures. Vertical lines represent twin boundaries in the $\rm LaAlO_3$ substrate}
    \label{Pattern}
\end{figure}

\section{Experimental}
\noindent
{\rm YBa$_2$Cu$_3$O$_{7-\delta}$/La$_{0.67}$Sr$_{0.33}$MnO$_3$} (YBCO/LSMO) bilayers were grown on 10 mm $\times$ 5 mm {\rm LaAlO$_3$}  single-crystalline substrates by multitarget high pressure sputtering from stoichiometric {\rm YBa$_2$Cu$_3$O$_{7-\delta}$} and {\rm  La$_{0.67}$Sr$_{0.33}$MnO$_3$} targets, at a substrate temperature of 770 $^\circ$ C and an oxygen pressure of 3 mbar. The deposition rates of {\rm YBa$_2$Cu$_3$O$_{7-\delta}$} and {\rm  La$_{0.67}$Sr$_{0.33}$MnO$_3$} were 2.6 and 1.6 nm per minute, respectively \cite{Przyslupski}. The resulting YBCO film thickness was 50 nm and LSMO films were 100 nm thick. Both YBCO and LSMO films were [001] oriented. The Curie temperature T$_C$ of the LSMO was about 360 K, while the YBCO superconducting transition temperature T$_c\,$ and the corresponding transition width $\Delta$T$_c$ were 89.3 K and 2.2 K, respectively.\\

\noindent
To enforce the motion of vortices at different angles to the stripe magnetic domains orientation a mask shown in Fig  \ref{Pattern}, containing six stripes separated by an angle of 18 degrees and covering in total the range of 90 degrees, was employed. To ensure a uniform magnetic domain pattern, the structures were fabricated on the preselected substrates, in which significantly large, larger than the entire six strips arrangement, zones of the uniform and regular twin boundaries were revealed through the optical microscopy. In such structures, the current in each stripe flows at a specific angle to the direction of stripe domains. The angle variation over 90$^\circ$ ensures covering the parallel and perpendicular flow direction, ie, the minimum and the maximum magnetic pinning range. The YBCO films with the metallic LSMO were patterned using UV photolithography and wet chemical etching in a HCl/H$_2$O/H$_2$O$_2$ (0.5\%/94.5\%/5\%) solution.\\

\begin{figure}[h]
    \centering
    \includegraphics[width=10cm]{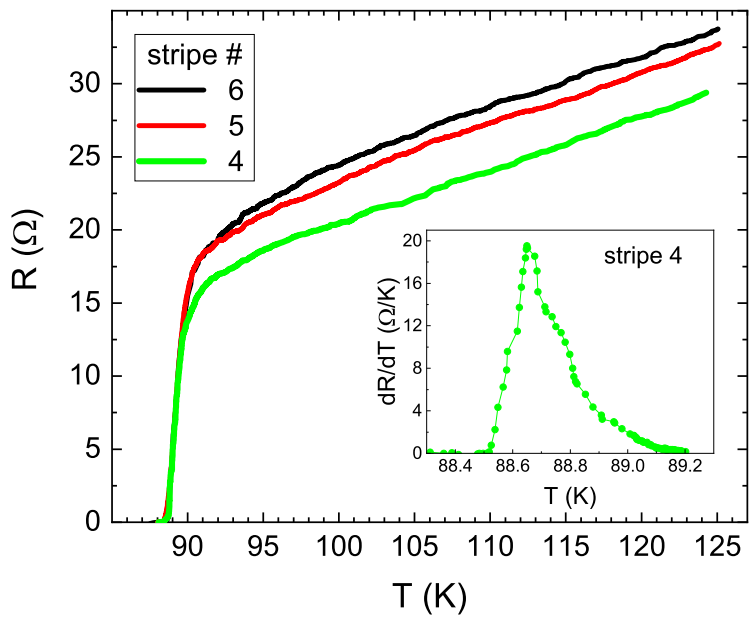}
    \caption{The temperature dependence of the resistance of three YBCO/LSMO stripes {\it vs} temperature. The inset shows the derivative $dR/dT$ for stripe \# 4 close to the superconducting transition}
    \label{RT_dRdT}
\end{figure}

\noindent
The critical temperature $T_c\,$ and the critical current $I_c\,$ of the stripes were measured in a standard dc four-point configuration, using a Janis cryostat with an open "pull mode" liquid nitrogen system.
The critical temperature was estimated by measuring the resistance {\it vs\,} temperature ($R(T)\,$
characteristics) at a dc bias current of 10 $\mu $A.
The critical current was determined from the current-voltage characteristics, using the field criterion of 5~$\times $~10$^2$ V/m.\\

\noindent
Alternatively, to obtain local values of the critical current density, we used direct visualization of the magnetic flux distribution in the samples by a  magneto-optical technique, which consists in obtaining an image of magnetic flux penetration into the superconductor using a Bi-doped yttrium-iron garnet indicator placed on the sample surface, rotating the polarization plane of the incident polarized light proportionally to the local magnetic field $B$. By calibrating the luminous intensity of magneto-optical images, we obtained a local induction map. The recorded profiles of the flux density $B$ taken across the samples are consistent with the critical state model of a superconductor  \cite{Bean, Zeldov}, which allows us to determine the local critical current density $j_c \approx (2/\mu_0)dB/dx$ \cite{Brandt}.
The results of two distinct techniques of critical current determination always fully coincided.\\

\noindent
The temperature in all transport measurements was measured by a calibrated diode thermometer and stabilized to $\pm$~0.1 K.\\

\section{Results and discussion}
\noindent
Examples of the resistance {\it vs} temperature $R(T)$ measurements for three stripes, each one oriented at a different angle to the domain structure, are presented in Fig \ref{RT_dRdT}. The inset shows the differential characteristics $dR(T)/dT\,$ of the $R(T)$ of stripe 4 around the normal-superconductive phase transition. The deducted transition temperature is $T_c  \approx$ 89.2 K, with the transition width $\Delta T_c  \approx $ 1.3 K. No significant variations of $T_c\,$ and $\Delta T_c\,$ between the different stripes were observed, although some variations in the normal state resistance are visible in Fig \ref{RT_dRdT}.\\

\begin{figure}[h]
    \centering
    \includegraphics[width=10cm]{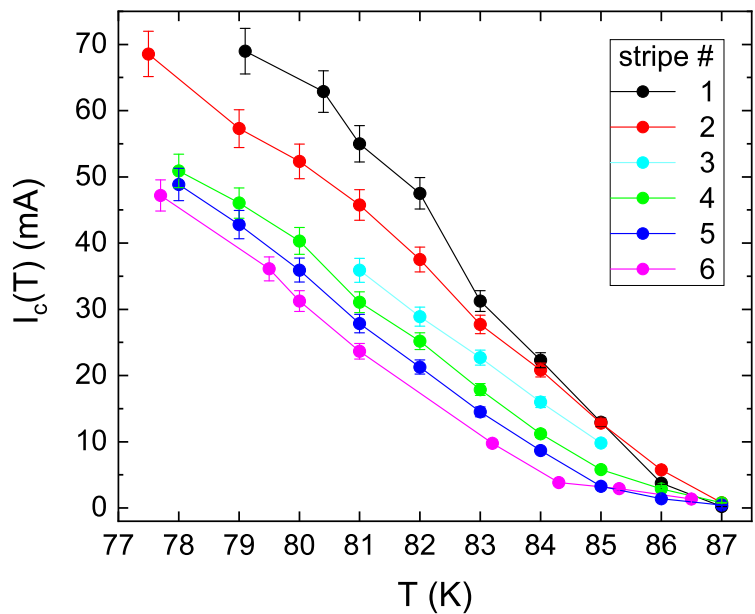}
    \caption{The temperature dependence of the critical currents for six YBCO/LSMO stripes directed at different angles $\alpha$ to the magnetic domain pattern}
    \label{Ic_T_6_stripes}
\end{figure}

\begin{figure}[h]
    \centering
    \includegraphics[width=10cm]{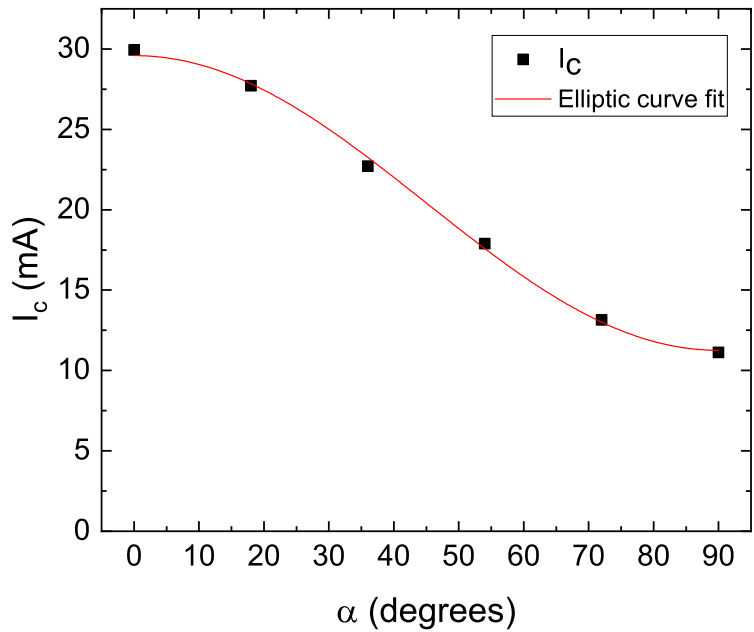}
    \caption{The angle dependence of the critical currents for six YBCO/LSMO stripes with different angle $\alpha$ for $T = 83~K$. The red line shows an elliptic best fit}
    \label{Ic_alpha_fit}
\end{figure}

\noindent
The transport properties of the stripe were found to depend systematically on the angle $\alpha\,$ between the vortex flow direction and the domain stripes. Measurements of the current-voltage characteristics ($I-V\,$) provided the critical vortex de-pinning current $I_c(T)\,$ as a function of temperature and vortex flow direction angle.
The critical current as a function of temperature is shown  in Fig \ref{Ic_T_6_stripes}, while the influence of the flow direction angle is illustrated in Fig \ref{Ic_alpha_fit}. It is apparent from Fig \ref{Ic_alpha_fit} that the critical current magnitude varies with the angle $\alpha$, which confirms the initial assumptions
that magnetic pinning magnitude varies with the  angle. Moreover, since the pinning force is highest for the orientation of the magnetic domains perpendicular to the direction of the vortex motion,  we can assume that stripe number 1 is directed parallel to the magnetic domain pattern, while stripe number 6  is perpendicular to the magnetic domain pattern.\\

\noindent
Vortices moving in each stripe can be considered as a superposition of motion in two orthogonal directions, parallel and perpendicular to the magnetic stripe directions. This corresponds to the motion in an anisotropic medium characterized by two different pinning magnitudes at perpendicular directions and the elliptic-like equation best describes the pinning force exerted on vortices \cite{Villard}.
The correctness of the motion description by the elliptic equation was confirmed by many authors, although some of them disputed the proposed physical mechanism of the anisotropy \cite{Yamasaki}.\\

\noindent
In our system, the anisotropy of vortex pinning is due to different strengths of the pinning force in the direction parallel and perpendicular to the magnetic domain orientation. Therefore, the $I_c\,$ variations with orientation angle should likewise follow the elliptic-like behavior. Following the  reasoning used by Villard {\it et al} \cite{Villard}, we assume that the considered YBCO/LSMO system,
can be described by a matrix relation between the  vectors of the electric field $\bf E\,$ and current density $\bf j\,$ in the stripes through a two-dimensional tensor of conductance with diagonal elements

\begin{equation}
\left( \matrix{
j_x \cr
j_y } \right)
 = \left( \matrix{
\sigma_x & 0 \cr
0 & \sigma_y } \right)
\left(\matrix{
E_x  \cr
E_y } \right) ,
\label{jx_jy}
\end{equation}

\noindent
where $x\,$ and $y\,$ are the directions parallel and perpendicular to the magnetic domains, respectively.\\

\noindent
If the applied electric field is rotated by an angle $\alpha $ in the $x,y\,$ plane, one obtains the elliptic formula for the magnitude of the current density $\bf j$\\

\begin{equation}
 |{\bf j} (\alpha ) | = \sqrt { \sigma_x^2 E_x^2\cos^2\alpha +  \sigma_y^2 E_y^2\sin^2\alpha }.
 \label{j_alpha}
\end{equation}

\noindent
Note, that equations (\ref{jx_jy}) and (\ref{j_alpha}) are applied here for a stripe in the superconducting dissipative state below $T_c$, ie for bias currents above the depinning critical current of the strip $I_c$.\\

\noindent
In the case of the critical current anisotropy in the YBCO film, induced here by the magnetic domains in a superconducting-ferromagnetic bilayer, one can write \cite{Safar}\\

\begin{equation}
\left( \matrix{
j_{cx} \cr
j_{cy} } \right)
 = \left( \matrix{
A_x & 0 \cr
0 & A_y } \right)
\left(\matrix{
I_{ax}  \cr
I_{ay}} \right) ,
\label{Ax_Ay}
\end{equation}

\noindent
where $j_{cx}\,$ and $j_{cy}\,$ are critical current components, $A_x\,$  and $A_y\,$ are diagonal elements of the critical current anisotropy tensor {\bf A}, and
$I_{ax}\,$ and $I_{ay}\,$ are components of the external applied current. Thus, rotating the applied current direction in the $x,y$-plane leads also to an elliptic behavior of the critical current density, as described by equation  (\ref{j_alpha}).\\

\noindent			
The dependence of the critical current $I_c$ on the orientation angle $\alpha$ at a selected temperature of 83 K is shown in Fig \ref{Ic_alpha_fit}. The dashed line shows the fit of the elliptic-like function to the $I_c (\alpha )$ data. The quality of the fit is very good, with $R^2$=0.998.\\

\begin{figure}[h]
    \centering
    \includegraphics[width=10cm]{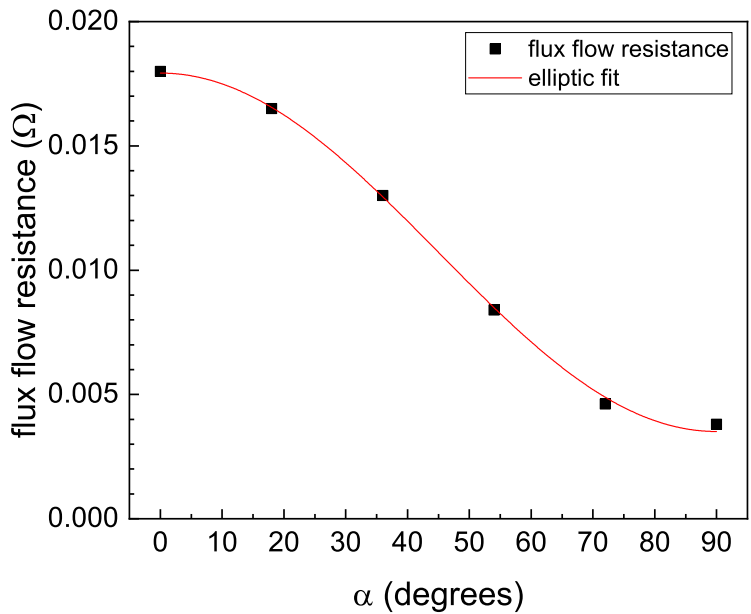}
    \caption{The angle dependence the flux-flow resistance for a YBCO/LSMO stripe {\it vs} angle $\alpha$ at $T = 83~K$. The solid line shows the best fit of the elliptic function to the data}
    \label{Rff_alpha_fit}
\end{figure}

\noindent
In the further analysis of the angular dependence, we concentrate on the dependence of the flux-flow resistance on the strip orientation angle $\alpha$. As pointed out above, with the varying  angle $\alpha$, the flux flow resistance should exhibit similar behavior to that of the critical current, as they are both similarly influenced by the anisotropic effect of magnetic domains. Figure \ref{Rff_alpha_fit} shows the flux flow resistance in the strips directed at different angles to the magnetic domains, together with the best fit of the elliptic function to the experimental data. Again, the fit quality is very good, $R^2$=0.998, confirming that the behavior of the flux flow resistance in the bilayer system can be indeed modeled similarly to that of the critical current.\\

\noindent
In addition, we have looked for the signatures of the coherence in guided vortex motion in bilayered samples by irradiating samples with microwave radiation and looking for the appearance of quasi-Josephson steps of the current-voltage characteristics of the YBCO stripes \cite{Yuzhelevski}. We did not observe any steps for vortices moving in the channels of easy vortex flow along the magnetic domain walls, ie at the angle $\alpha = 0^o$. This is not surprising, because the density of current-induced vortices is relatively low and consequently, the vortex-vortex interactions are weak, not sufficient to enforce coherence in their motion. However, we have observed Shapiro-like steps in the $I-V\,$ characteristics of the strips directed perpendicular to the domain walls, with the first step voltage position $V_1\,$ given by the Josephson frequency-voltage relationship $V_1=\frac{h}{2e}f$, where $h\,$ is the Planck`s constant, $e\,$ electron charge and $f\,$ the irradiation frequency. In fact, vortex motion in such strips is equivalent to the vortex flow in a periodic pinning potential, known to result in pronounced manifestations of the quasi-Josephson effects \cite{Yuzhelevski}. The microwave-induced step structures were also observed on the $I-V$ characteristics of two adjacent strips, located at angles close to $\alpha = 90^o$, nevertheless steps structures in these strips were much weaker and blurred.\\

\begin{figure}[h]
    \centering
    \includegraphics[width=10cm]{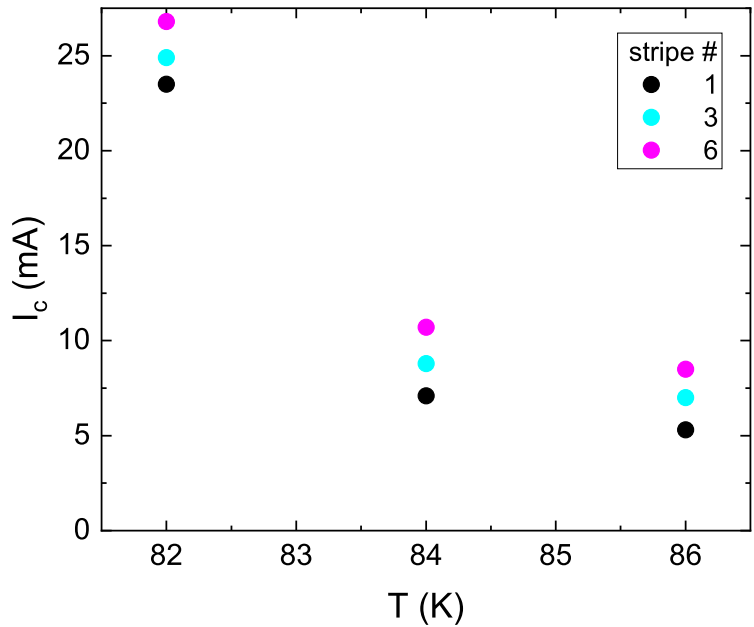}
    \caption{The temperature dependence of the critical currents for three YBCO stripes
    directed at three different angles $\alpha$ to the dominating direction of twin boundaries}
    \label{Ic_only_YBCO}
\end{figure}

\noindent
The employed YBCO/LMSO bilayers are deposited on ${\rm LaAlO_3}$  substrates which naturally trigger the formation of twin domains and twin boundaries as a stress relief mechanism during the sample deposition \cite{Koren}. Such twin defects alone may give rise to significant pinning potential in the YBCO film, even in the absence of magnetic interactions. To verify what part of the pinning anisotropy in our experiments is due to twin boundaries alone, we fabricated and measured test samples of bare YBCO thin film deposited on the ${\rm LaAlO_3}$ substrate without underlying LMSO layer. We have found that twin boundaries produce weaker anisotropy effects than those enhanced by the magnetic domain interactions. Figure \ref{Ic_only_YBCO} shows the results of the critical current measurements performed with a test sample, YBCO alone on a twinned substrate, for strips directed close to 0, 36, and 90 degrees to the  direction of the twin pattern. By comparing these results with data obtained using YBCO/LSMO bilayers one concludes that magnetic interactions enhance the pinning anisotropy by a factor close to 2. Thus, we conclude that the dominant source of vortex pinning in our system is indeed magnetic domain structure.\\

\subsection{Conclusions}
\noindent
In conclusion, we have investigated the influence of the angles $\alpha\,$ between magnetic domain orientation and vortex motion direction on the vortex pinning and consequently on the motion of magnetic flux vortices in YBCO/LSMO bilayers. We experimentally established that vortex interaction with magnetic domains enhances and doubles the pinning anisotropy of the system. The anisotropic pinning leads to an elliptic-like behavior of both the critical current and flux flow resistivity. Motion of vortices in the direction perpendicular to the magnetic domain walls has a coherent character and leads to the appearance of the quasi-Josephson current steps on the $I-V$ characteristics. We believe that our results may contribute to the design of  future cryoelectronics devices based on guided vortex motion.

\ack
Transport measurements presented in this paper were executed by Dr Guy Bareli.

\end{document}